# Marine animal behaviour: neglecting ocean currents can lead us up the wrong track

Philippe Gaspar[(1)*], Jean-Yves Georges[(2)*], Arnaud Lenoble[(1)], Sandra Ferraroli[(2,3)], Sabrina Fossette[(2,3)] and Yvon Le Maho[(2)]

[(1)] Collecte Localisation Satellites, Direction Océanographie Spatiale, 8-10 rue Hermès, 31520 Ramonville, France

[(2)] Centre National de la Recherche Scientifique, Institut Pluridisciplinaire Hubert Curien, Département Ecologie, Physiologie et Ethologie, Unité Mixte de Recherche 7178 CNRS – Université Louis Pasteur, 23 rue Becquerel, 67087 Strasbourg, France

[(3)] Université Louis Pasteur, 4 rue Blaise Pascal, 67070 Strasbourg, France

Corresponding authors:

**Philippe Gaspar***

Phone: + 33 561 394 781

Fax: +33 561 393 782

philippe.gaspar@cls.fr

**Jean-Yves Georges***

+33 388 106 947

+33 388 106 906

jean-yves.georges@c-strasbourg.fr

**\* PG and JYG contributed equally to this work**

**Abstract**

**Tracks of marine animals in the wild, now increasingly acquired by electronic tagging of individuals, are of prime interest to identify habitats and high-risk areas, but also to gain detailed information about the behaviour of these animals. Using recent satellite-derived current estimates and leatherback turtle (*Dermochelys coriacea*) tracking data, we demonstrate that oceanic currents, usually neglected when analysing tracking data, can substantially distort the observed trajectories. Consequently, this will affect several important results deduced from the analysis of tracking data, such as the evaluation of the orientation skills and the energy budget of animals or the identification of foraging areas. We conclude that currents should be systematically taken into account to ensure the unbiased interpretation of tracking data which now play a major role in marine conservation biology.**

Key index words: *biologging*, travelling *versus* foraging behaviour, impact of ocean currents, marine animal behaviour, satellite oceanography, wildlife tracking.

## 1. Introduction

Understanding how marine animals utilize the oceanic environment and its constraints is crucial for the development of sound management strategies for marine ecosystems that are threatened by climate change (Parmesan & Yohe 2003; Schmittner 2005) and direct anthropogenic pressure (Myers & Worm 2003; Lewison *et al.* 2004a). This could not be achieved without the detailed observation of free-ranging organisms, which is now possible through the electronic tagging of individuals (Block 2005). In the last few years, this *biologging* approach has been widely used for various marine species, greatly enhancing our knowledge of physiology, habitat use and movements of long-distance migrants (McConnell *et al.* 1992; Bost *et al.* 1997; Lutcavage *et al.* 1999; Costa & Sinervo 2004; Block *et al.* 2005). *Biologging* also provided us with unique observations, which are helpful in developing effective marine ecosystem conservation measures. In particular, electronic tracking provided evidence that the foraging hot spots frequented by pelagic fish and their associated fishing fleets are also exploited by endangered species, such as sea turtles (Spotila *et al.* 1996). Hence, the probability of incidentally catching endangered species is greatly increased in these areas (Ferraroli *et al.* 2004; Hays *et al.* 2004a; Lewison *et al.* 2004b). Accurately locating these zones and understanding how animals use these critical areas is indispensable if we want to define how, where, and when fishery management procedures must be applied, so that we can ensure the sustainable exploitation of commercial species while also minimizing the by-catch of endangered species (Lewison *et al.* 2004a; 2004b).

Identifying when and where marine animals forage is therefore crucial and such information can be deduced from the recorded trajectories of tagged animals (Sibert *et al.* 1999; Benhamou 2004; Newlands *et al.* 2004, Gutenkunst *et al.* 2006). Other information that can be deduced from track analyses includes the identification of migration corridors (Morreale *et al.* 1996), dispersion patterns (e.g. Block *et al.* 2005), environmental preferences (e.g. Polovina *et al.* 2004) and navigation strategies (e.g. Akesson *et al.* 2003).

While the influence of ocean currents on recorded trajectories has been detected and qualitatively assessed (Luschi *et al.* 1998; Hays et al., 1999; Polovina *et al.* 2000; Luschi *et al.* 2003a, 2003b), their impact on the results of track analyses has never been quantitatively assessed. The lack of detailed current estimates along the trajectories of aquatic species has prevented such an assessment, while the effects of wind on migrating birds are well documented (e.g. Richardson 1990; Weimerskirch et al. 2000). Recent progress in satellite oceanography (Rio & Hernandez 2003; 2004) now allows the synoptic estimation of surface currents. We use these current estimates here to conduct a first quantitative evaluation of the impact of oceanic currents on the trajectories of marine animals and investigate the consequences for the results deduced from tracking analysis.

## 2. Methods

*(a) Separating swimming and drifting*

The trajectories of marine animals reflect the combined effects of the animal's voluntary motion (swimming) and its transportation by oceanic currents (drift). More precisely, trajectories are time series of the animal's location [$\mathbf{X}(t_0)$, $\mathbf{X}(t_1)$ ..., $\mathbf{X}(t_N)$] at the sea surface. The observed velocity of the animal over ground ($\mathbf{V_g}$) is the time derivative of $\mathbf{X}$. In practice, $\mathbf{V_g}$ is estimated by computing the distances, in the x and y directions, between two consecutive locations and then dividing these distances by the time elapsed. This velocity is the sum of the animal's swimming velocity ($\mathbf{V_s}$) plus the velocity of the fluid in which the animal moves, i.e. the velocity of the current ($\mathbf{V_c}$):

$$\mathbf{V_g} = \mathbf{V_s} + \mathbf{V_c} \tag{1}$$

Tracks are thus simple linear combinations of the animal's own motion and the motion of the surrounding fluid:

$$\mathbf{X}(t) = \mathbf{X}(t_o) + \int_{to}^{t} \mathbf{V_g}(t)\,dt = \mathbf{X}(t_o) + \int_{to}^{t} \mathbf{V_s}(t)\,dt + \int_{to}^{t} \mathbf{V_c}(t)\,dt \tag{2}$$

Depending on whether the animal, or the ocean, is quiet or active, the relative importance of these two components on the observed motion can be highly variable and affect most results regarding the animal's behaviour inferred from track analyses.

*(b) Processing satellite tracking data*

To quantify the impact of ocean currents in a wide range of oceanic conditions, we analyze here a very long (11 635 km, 295 days) trajectory of an Argos-tracked female leatherback turtle (*Dermochelys coriacea*), hereafter referred to as T8 (Figure 1). T8 left French Guyana, South America in June 29, 2000 and circulated through most of the North Atlantic Ocean, crossing major current systems and quiet oceanic areas, before transmission stopped on April 20, 2001.

The trajectory of T8 was edited as follows. All Argos locations implying an apparent velocity above 10 km/h were discarded (Eckert 2002). The track was then smoothed and re-sampled at 3-hour intervals using simple local linear regression with a time window of 2 days. Such smoothing is needed to filter out most of the Argos location error which acts as high-frequency noise that would subsequently be amplified in the velocity derivation process. The associated re-sampling is a standard procedure allowing the computation of homogeneous velocities ($V_g$) over constant time intervals. With 3-hour intervals, the mean distance between successive re-sampled positions is close to 5 km. This spatial resolution provides a more than sufficient sampling of the mesoscale variations of the ocean current field (see next section).

*(c) Estimating currents from satellite observations*

Strictly speaking, the velocity of the current ($V_c$) should be estimated at the depth of the tracked animal. This is no simple task as currents are not easily monitored over vast oceanic areas, except for surface currents which can now be estimated using satellite observations (e.g. Ducet et al., 2000). Fortunately, leatherback turtles mostly dive in the near-surface epipelagic zone (Hays *et al.* 2004b), where the current velocity generally differs little from its surface value (see discussion below). Satellite-derived surface current estimates can thus be used to approximate $V_c$. They are obtained as the sum of the mean ($V_m$) and anomaly ($V_a$) of the surface geostrophic current plus the surface Ekman current ($V_e$):

$$V_c = V_m + V_a + V_e \qquad (3)$$

The mean geostrophic velocity is provided by Rio and Hernandez (2004) on a regular 1° x 1° grid. The geostrophic relation is used to deduce $V_a$ from gridded fields of sea level anomalies (SLA) measured by the radar altimeters onboard the TOPEX-POSEIDON and ERS-2 satellites. Weekly SLA fields are obtained from AVISO (http://www.aviso.oceanobs.com) on a 1/3° x 1/3° Mercator grid. The Ekman component of the current ($V_e$) is computed as a function of the surface wind stress using the Rio and Hernandez model (2003). Daily wind stresses, derived from QuickSCAT scatterometer measurements, are obtained from CERSAT (http://www.ifremer.fr/cersat) on a regular 25 km x 25 km grid. Then, based on these gridded velocity fields, the three components of the surface current are linearly interpolated (in space and time) at each position of the re-sampled track.

This surface velocity reconstruction technique was recently evaluated by Pascual et al. (2006) who compared so-estimated velocities to a large set (over 600 000 measurements) of surface velocity observations from the Global Drifter Program (http://www.aoml.noaa.gov/phod/dac). Velocity estimates prove to be essentially unbiased with a mean error below 1 cm/s for both the meridional and the zonal component of the velocity vector. In energetic areas (with root-mean-square velocities above 20 cm/s), estimated velocities explain 73.3 % of the drifter zonal velocity variance and 66.4 % of the drifter meridional velocity variance. Most of the unexplained variance appears to lie in high-frequency signals, not resolved by altimetric observations (Le Traon & Dibarboure, 2002). The comparison with drifter observations excluded shallow coastal areas, where significant deviations from geostrophic equilibrium are known to occur, and the Equatorial band, where both the geostrophic and Ekman approximations break down. In our case, these limitations only impact the first 4 days of tracking after which T8 leaves the Guiana shelf and enters the open waters of the North Atlantic Ocean. The reconstructed currents thus provide realistic surface velocity estimates along nearly all of the T8 track. Still, as pointed out above, surface velocity estimates are only used as a proxy for the current velocity at the depth of the tracked animal. This approximation is justified here as recent observations of the open-ocean diving behaviour of leatherback turtles indicate that adult leatherbacks, like T8, generally spend over 50 % of their time near the surface (typically between 0 and 50 m) and rarely dive below 200 m (Hays et al, 2004b; James et al., 2005). These animals thus mostly occupy the upper oceanic layers where the vertical variations of the current are essentially the vertical variations of its Ekman component. This relatively small current component (mean velocity of the Ekman current below 3 cm/s for a mean total current velocity of 19 cm/s along T8 track) is maximum at the surface and decreases with depth. An animal regularly diving from the surface to modest depths shall thus experience an average current speed differing from the surface current value by only a few cm/s. More important deviations from the surface current will only be experienced in rare circumstances when the turtle dives down to several hundred meters (Hays et al., 2004b) below the core of major surface currents or mesoscale features.

*(d) Defining a current-corrected track*

At first sight, the relation between the apparent movement of T8 and the current direction (Figure 1) is not at all obvious. It is better analyzed comparing (Figure 2) the observed track with the current-corrected track $\mathbf{X_{cc}}(t)$, that is the trajectory that the animal would have followed in a motionless ocean, swimming exactly the way she swam:

$$\mathbf{X_{cc}}(t) = \mathbf{X}(t_o) + \int_{t_o}^{t} \mathbf{V_s}(t)\, dt \qquad (4)$$

In practice, this current-corrected track is readily computed using the available tracking and current data:

$$\mathbf{X_{cc}}(t_o + n\Delta t) = \mathbf{X}(t_o) + \sum_{i=0}^{n-1} \Delta t\, [\mathbf{V_g} - \mathbf{V_c}](t_o + i\Delta t), \text{ for } n = 1, N \qquad (5)$$

where $\Delta t$ is the period at which the observed track has been re-sampled and N is the total number of locations in this track.

**3. Results**

*(a) New evidence for strong compass sense*

Between departure (A) and 34°N, over more than 3250 km, the observed track is remarkably straight, and the current-corrected track is even slightly straighter. Then between 34°N and milestone B (another 1000 km), the observed trajectory displays several marked heading changes while the current intensifies as T8 crosses the Gulf Stream system. Interestingly, unlike the observed track, the current-corrected track (Figure 2) remains remarkably straight in this highly dynamic area. This reveals an interesting navigation strategy in which T8 does not change her swimming direction to compensate for the current drift but, on the contrary, maintains a steady heading while crossing strong currents pushing at cross-angles. This behaviour, not detectable without current correction, is an additional proof of the strong compass sense previously detected in leatherback turtles (Lohmann & Lohmann 1993).

*(b) Swimming and drifting: impact on energy budget*

After a period of apparent erratic motion (between B and C), T8 started moving East in the frontal area between the subtropical and subpolar gyres (between C and D) where she is pushed by the powerful flow of the Gulf Stream Extension and then the North Atlantic Current with estimated velocities often above 0.5 m/s, peaking at 0.95 m/s. Current impact on the track is remarkable: the current-corrected track reveals that T8 has little active motion towards the East. The distance actually swam by T8 between C and D (length of the current-corrected track) is only 1196 km while the distance travelled is 2144 km. Thus, almost half of the observed displacement is due to the current drift. Since the work done by the animal is essentially proportional to the distance swum, any energy budget using the distance computed along the observed track would be grossly wrong. Segment C-D is obviously the track segment where the current impact is the largest. Over the other segments, the difference between the current-corrected track length and the observed track length remains between 5 and 20 % of the later.

*(c) Navigation strategy in the presence of strong currents*

In the last recorded part of the trajectory (D to G), T8 appears to progressively come back to the Western part of the Atlantic basin, with the conspicuous exception of the E-F segment. But, the current-corrected track (Figure 2) reveals that, even if apparent motion from E to F is towards East, T8 actually kept swimming towards West between these two points. More precisely, computed velocities indicate that, while T8 was swimming at 0.11 m/s (mean of the westward component of her swimming velocity), the Azores current, centred around 35°N, pushed her faster to the East (mean of the eastward component of the current velocity: 0.20 m/s). The convoluted shape of the E-F segment (Figure 1) indicates that T8 was actually foraging, while slowly swimming against the current, in the very productive Azores front. After this feeding event, the T8 trajectory becomes straighter as she steadily heads North-West (right after F). By doing so T8 gets out of the narrow core of the Azores current and starts actually moving West again in a region, where her motion is no longer opposed by a strong eastward flow. She then heads to the South, encountering the Azores current again. This time, she crosses it completely and then keeps moving to the West. Further work is needed to understand the exact navigation mechanisms used to shape such a complex trajectory but our analysis shows that this can only be done with detailed knowledge of the currents.

*(d) Diagnosing travelling or foraging behaviour in the presence of currents*

Going one step further, new current data allow us to revisit more elaborate track analyses aiming at the identification of the two main movement behaviours, i.e. travelling versus food prospecting. These two motion types are characterized by long-ranged directed displacements and erratic motions, respectively. Based on this simple characterization, several techniques (e.g. Benhamou 2004; Newlands *et al.* 2004) have been developed to automatically identify track segments where either travelling (directed motion) or foraging (erratic motion) dominates. Such analyses are extremely useful when no other behavioural information is available. They provide important information on feeding ground locations, migration areas, switching frequency between travelling and foraging behaviour (Newlands *et al.* 2004). Track analyses are also useful when combined with complementary biologging measurements (e.g. dive profiles) to obtain more precise descriptions of the animal's movement and behaviour in relation to its environment (e.g. Georges et al., 2000; LeBoeuf et al., 2000; Hays et al., 2004b, James et al., 2005; Sale et al., 2006).

It is plain that this movement characterization (erratic or directed) concerns the animal's own motion ($V_s$) and not the combined motion of the animal and surrounding fluid ($V_s+V_c$). In other words, such track analyses should be performed on current-corrected tracks, not on observed tracks as usually done.

To show the differences between the two approaches we analyzed the impact of the current correction on the straightness index S of the T8 track. This index, defined by the ratio of the beeline distance between two points on the track and the actual length of the track between these two points (Batschelet 1981), is often used to distinguish between travelling and foraging behaviours in track analyses (Benhamou 2004). If the animal travels straight from one point to the other, S=1, whereas S → 0 in the case of erratic motion. We computed the value of S to the middle of 4-day long periods for both the actual and current-corrected tracks. This period of 4 days was chosen to detect travelling or foraging events extending over several days as the original sampling frequency of data, and their subsequent smoothing, prevents detailed analysis at shorter periods. Both the original and current-corrected straightness indices showed a bimodal distribution, with a S=0.8 threshold value separating both modes. Accordingly, we used S=0.8 for separating travelling (S>0.8) from foraging (S<0.8) behaviour, for both the original and current-corrected straightness indices (Figure 3). Figure 3 shows that current correction does not modify the "travelling diagnosis" on segments A-B, D-E and F-G, nor the "foraging diagnosis" on B-C and F-G. But on segment C-D, the uncorrected index points at a dominant travelling behaviour (S>0.8 during 51 days of this 61-day period), whereas the current-corrected value of S clearly indicates a foraging behaviour, except for the last 7 days of that period. The difference in the diagnosis is huge as the current-corrected index suggests that, between C and D, T8 foraged during 54 days along a 1807-km track segment (Figure 4). On the contrary, the uncorrected index only points at a 10-day, 191-km long, foraging episode. The diagnosis of a much longer foraging period obtained with the current-correction is otherwise supported by the fact that the concerned track segment between C and D lies in the transition zone between the subtropical and subpolar gyres where food abounds (Polovina *et al.* 2001). In addition, T8 was particularly slow along that C-D segment (mean swimming velocity: 0.26 m/s), suggesting that she was indeed searching and/or processing food.

**4. Discussion**

T8 track is remarkable as it displays various types of animal movement patterns occurring in various ocean regions with widely different dynamics. Analysis of over ten other postnesting trajectories of leatherback turtles tracked from French Guyana (Ferraroli et al., 2004) provide similar results, even if no single other trajectory is as complete and as informative as T8 track. In all trajectories, currents prove to commonly add sinuosity to long-range directed track segments confirming that leatherback turtles tend to maintain a more stable heading than indicated by the observed tracks. Their compass sense is truly remarkable, even in the presence of strong currents. The other tracks also provide examples of (probably) foraging turtles either swimming with the current (like on the C-D segment), or actively swimming against it (like on the E-F segment). The reason for such different navigation strategies relative to the current needs to be further analysed with support from complementary diving data.

Altogether, our results reveal that currents have a highly variable but rarely negligible impact on marine animals' tracks. Ignoring current effects can thus be misleading when trying to infer animal behaviour from tracking data. Additionally, precise information on the currents significantly modifies, and moreover improves, our interpretation of observed trajectories. We thus conclude that currents should be systematically taken into account to warrant unbiased analysis of marine animals' tracking data. Surface currents, derived from satellite observations, are readily available to analyze tracks from (mostly) epipelagic animals. For deeper-diving animals, realistic current estimates at all depths should soon be available from operational global ocean models.

Marine animals faster than leatherback turtles will clearly be less impacted by currents but oceanographic information will remain important for interpreting their actual behaviour. An accurate determination of the currents will indeed be needed to understand how travelling animals shape their trajectories as a function of currents. This might help unravel some of the remaining mysteries of animal navigation.

Currents will be even more important to analyze foraging behaviour. Indeed, foraging typically occurs in rich, dynamically active, areas where currents tend to be faster while feeding animals tend to slow down. In such areas the balance between oceanic movements and animal motion will be subtle. Current correction will thus be critically needed to properly assess foraging habitat and the time budget of individuals. This shall lead to a more accurate estimation of food requirements, and ultimately to a better quantification of the way animal populations impact on trophic resources and respond to changes in food availability, throughout marine ecosystems.

**Acknowledgments**

We thank the customary chiefs and inhabitants of Awala-Yalimapo (French Guyana) and the Amana Natural Reserve for their support in the field, and Ph. Schaeffer for assistance with data visualization. This study was carried out under CNRS institutional license (B67 482 18) and adhered to the legal requirements of the country in which the work was carried out, and all institutional guidelines.

**References**

Akesson, S., Broderick, A.C., Godley, B.J., Lushi, P., Papi, F. & Hays, G.C. 2003 Navigation by green turtles; which strategy do displaced adults use to find Ascension Island? Oïkos 103 (2): 363-372.

Batschelet, E. 1981 *Circular statistics in Biology*, London, Academic Press.

Benhamou, S. 2004 How to reliably estimate the tortuosity of an animal's path: straightness, sinuosity or fractal dimension? *J. Theor. Biol.* **229**, 209-220.

Block, B. 2005 Physiological ecology in the 21$^{st}$ century: advancements in biologging science. *Integr. Comp. Biol.* **43**, 305-320.

Block, B.A. Teo, S.L.H, Walli, A., Boustany, A., Stokesbury, M.J.W., Farwell, C.J., Weng, K.C., Dewar, H. & Williams, T.D. 2005 Electronic tagging and population structure of Atlantic bluefin tuna. *Nature* **434**, 1121-1127.

Bost, C.A., Georges, J-Y, Guinet, C., Cherel, Y., Pütz, K., Charrassin, J.B., Handrich, Y., Zorn, T., Lage, J. & Le Maho, Y. 1997 Foraging habitat and food intake of satellite-tracked King penguins during the austral summer at Crozet Archipelago. *Mar. Ecol. Prog. Ser.* **150**, 21-33.

Costa, D.P. & Sinervo, B. 2004 Field physiology: physiological insights from animals in nature. *Annu. Rev. Physiol.* **66**, 209-238.

Ducet, N., Le Traon, P.Y. & Reverdin, G. 2000 Global high-resolution mapping of ocean circulation from TOPEX/Poseidon and ERS-1 and -2. *J. Geophys. Res.* **105**, 19477-19498.

Eckert, S.A. 2002 Swim speed and movement patterns of gravid leatherback sea turtles (*Dermochelys coriacea*) at St Croix, US Virgin Island. *J. Exp. Biol.* **205**, 3689-3697.

Ferraroli, S., Georges, JY., Gaspar, P. & Le Maho, Y. 2004 Where leatherback turtles meet fisheries. *Nature* **429**, 521-522.

Georges, J.Y, Bonadonna, F. & Guinet, C. (2000) Foraging habitat and diving activity of lactating subantarctic fur seals in relation to sea surface temperatures on Amsterdam Island. *Mar Ecol. Prog. Ser.* **196** : 291-304

Gutenkunst, R., Newlands, N., Lutcavage, M. & Edelstein-Keshet, L. 2005 Inferring resource distributions from Atlantic bluefin tuna movements: an analysis based on net displacement an length of track, submitted to *J. Theor. Biol*.

Hays, G. C., Houghton, J.D.R., Isaac, C., King, R.S., Lloyd, C. & Lovell, P. 2004b First records of oceanic dive profiles for leatherback turtles (*Dermochelys coriacea*) indicate behavioural plasticity associated with long distance migration. *Anim. Behav.* **67**, 733-743.

Hays, G.C., Houghton, J.D.R. & Myers, A.E. 2004a Pan-Atlantic leatherback turtle movements. *Nature* **429**, 522.

Hays, G.C., Luschi, P., Papi, F., del Seppia, C. & Marsh, R. 1999 Changes in behaviour during the internesting period and postnesting migration for Ascension Island green turtles. *Mar. Ecol. Progr. Ser.* **189**, 263-273.

James, M.C., Myers, R.A. & Ottensmeyer, C.A. 2005 Behaviour of leatherback sea turtles, Dermochelys coriacea, during the migratory cycle. *Proc. R. Soc. B* **272,** 1547-1555.

Le Traon, P.Y., & Dibarboure, G. 2002 Velocity mapping capabilities of present and future altimeter missions : The role of high frequency signals. *J. Atmos. Oceanic Technol.* **19,** 2077-2088.

LeBoeuf, B.J., Crocker, D.E., Costa, D.P., Blackwell, S.B., Webb, P.M. & Houser D.S. 2000 Foraging ecology of northern elephant seals. *Ecological Monographs* 70: 353-382

Lewison, R.L., Crowder, L.B., Read, A.J. & Freeman, S.A. 2004a Understanding impacts of fisheries bycatch on marine megafauna. *Trends in Ecology and Evolution* **19**, 598-604.

Lewison, R.L., Freeman, S.A. & Crowder, L.B. 2004b Quantifying the effects of fisheries on threatened species: the impact of pelagic longlines on loggerhead and leatherback sea turtles. *Ecology Letters* **7**, 221-231.

Lohmann, K. J. & Lohmann, C. M. F. 1993 A light-independent magnetic compass in the leatherback sea turtle. Biol. Bull., **185**, 149-151

Luschi, P., Hays, G.C. & Papi, F. 2003a A review of long-distance movements by marine turtles, and the possible role of ocean currents. *Oikos* **103**, 293-302.

Luschi, P., Hays, G.C., Del Seppia, C., Marsh, R. & Papi, F. 1998 The navigational feats of green sea turtles migrating from Ascension Island investigated by satellite telemetry. *Proc. R. Soc. B* **265**, 2279-2284.

Luschi, P., Sale, A., Mencacci, R., Hugues, G. R., Lutjeharms, J. R. E. & Papi, F. 2003b Current transport of leatherback sea turtles (*Dermochelys coriacea*) in the ocean. *Proc. R. Soc. London. B* **270**, S129-S132 DOI 10.1098/rsbl.2003.0036

Lutcavage, M.E., Brill, R.W., Skomal, G.B., Chase, B.C. & Howey, P.W. 1999 Results of pop-up satellite tagging of spawning size class fish in the Gulf of Maine: do North Atlantic bluefin tuna span in the mid-Atlantic? *Can. J. Fish. Aquat. Sci.* **56**, 173-177.

McConnell, B.J., Chambers, C. & Fedak, M.A. 1992 Foraging ecology of southern elephant seals in relation to the bathymetry and productivity of the Southern Ocean. *Antarct. Sci.* **4**, 393-398.

Morreale, S.J., Standore, E.A., Spotila, J.R. & Paladino, F.V. 1996 Migration corridor for sea turtles. *Nature* **348**, 319-320.

Myers, R.A. & Worm, B. 2003 Rapid worldwide depletion of predatory fish communities. *Nature* **423**, 280-283.

Newlands, N.K., Lutcavage, M.E. & Picker, T.J. 2004 Analysis of foraging movements of Atlantic bluefin tuna (*Thunnus thynnus*): individuals switch between two modes of search behaviour. *Popul. Ecol.* **46**, 39-53, doi:10.1007/sl10144-004-0169-9.

Parmesan, C. & Yohe, H. 2003 A globally coherent fingerprint of climate change impacts across natural systems. *Nature* **421**, 37-42.

Pascual, A., Faugère, Y., Larnicol, G., & Le Traon, P.Y. 2006 Improved description of the ocean mesoscale variability by combining four satellite altimeter missions. *Geophys. Res. Let.* **33** L02611, doi:10.1029/2005GL024633

Polovina, J. J., Kobayashi, D.R., Parker, D.M., Seki, M.P. & Balazs, G.H. 2000 Turtles on the edge: movement of loggerhead turtles (*Caretta caretta*) along oceanic fronts, spanning longline fishing grounds in the central North Pacific, 1997-1998. *Fish. Oceanogr.* **9**, 71-82.

Polovina, J.J., Howell, E., Kobayashi,D.R. & Seki, M.P. 2001 The transition zone chlorophyll front, a dynamic global feature defining migration and forage habitat for marine resources. *Progr. Oceanogr.* **49**, 469-483.

Polovina, J.J., Balazs, G.H., Howell, E.A., Parker, D.M., Seki, M.P. & Dutton, P.H. 2004 Forage and migration habitat of loggerhead (*Caretta caretta*) and olve ridley (*Lepidochelys olivacea*) sea turtles in the central North Pacific Ocean. *Fish. Oceanogr.*, **13**, 36-51

Richardson W.J. 1990 Wind and orientation of migrating birds: A review. *Cellular and Molecular Life Sciences* **46 (4),** 416-425

Rio, M.-H. & Hernandez, F. 2003 High-frequency response of wind-driven currents measured by drifting buoys and altimetry over the world ocean. *J. Geophys. Res.* **108**, 3283, doi:10.1029/2002JC001655.

Rio, M.-H. & Hernandez, F. 2004 A mean dynamic topography computed over the world ocean from altimetry, in situ measurements, and a geoid model. *J. Geophys. Res.* **109**, C12032, doi:10.1029/2003JC002226.

Sale, A., Luschi, P., Mencacci, R., Lambardi, P., Hugues, G.R., Hays, G.C., Benvenuti, S. & Papi, F. 2006 Long-term monitoring of leatherback turtle diving behaviour during oceanic movements. *J. Exp. Mar. Biol. Ecol.* **328**, 197-210

Schmittner, A. 2005 Decline of the marine ecosystem caused by a reduction in the Atlantic overturning circulation. *Nature* **434**, 628-633.

Sibert, J.R., Hampton, J., Fournier, D.A. & Bills, P.J. 1999. An advection-diffusion-reaction model for the estimation of fish movement parameters from tagging data, with application to skipjack tuna (*Katsuwonus pelamis*). *Can. J. Fish. Aquat. Sci.*, **56,** 925-938

Spotila, J.R., Dunham, A.E., Leslie, A.J., Steyermark, A.C., Plotkin, P.T. & Paladino, F.V. 1996 Worldwide population decline of *Dermochelys coriacea*: are leatherback turtles going extinct? *Chelon. Cons. Biol.* **2**, 209-222.

Weimerskirch, H., Guionnet, T., Martin, J., Shaffer, S.A. & Costa, D.P. 2000. Fast and fuel efficient? Optimal use of wind by flying albatrosses. *Proc. Biol. Sci.* 267:1869-74.

**Figure captions**

**Figure 1:** Trajectory of Argos-tracked leatherback turtle T8 (red line) with superimposed surface current vectors (green). This track is re-sampled at 3-hourly intervals and a current vector is plotted every 12 hours. Each vector has its origin on the track. Black dots labelled A to G are used for track segmentation. T8 reaches these milestones at the following dates: A(departure): 29/06/2000, B: 10/09/2000, C: 14/10/2000, D: 14/12/2000, E: 17/01/2001, F: 22/02/2001, G (end of data transmission): 20/04/2001. A zoom showing details of the E-F segment is inserted.

**Figure 2:** Comparison of the observed (red) and current-corrected (blue) track of T8. Milestones are positioned on the current-corrected track at the same dates as on the observed track.. A zoom showing details of the B-C segment is inserted. Note that the current-corrected track displays the animal's own motion ($V_s$) integrated in time but, taken alone, a position along that track bears no direct interpretation (and could very well be on the continent!).

**Figure 3:** Straightness index (S) computed along the observed (red) and current-corrected (blue) tracks. S is computed over 4-day long segments and the computed value is attributed to the central time of each segment. This yields values of S every 3 hours, except for the first and last two days of the tracking period. Travelling is diagnosed for S>0.8 and foraging for S<0.8.

**Figure 4:** Localisation of the travelling (red) and foraging (green) segments on the observed track, based on the value of the current-corrected straightness index. All fixes corresponding to S> 0.8 are plotted as red dots. Fixes with S<0.8 are plotted as green dots.

**Short title for page heading** : Ocean currents and tracks of marine animals

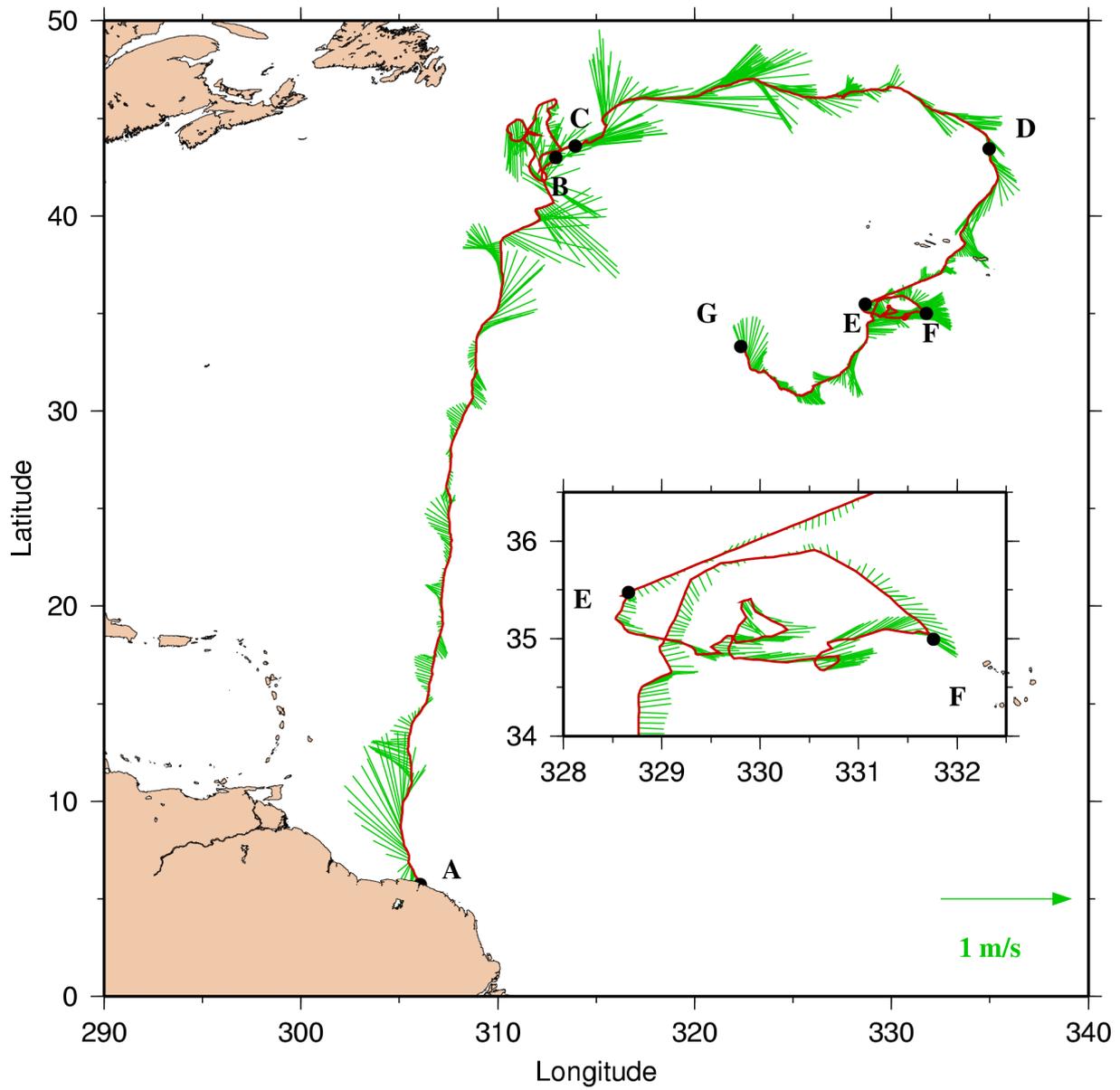

**Fig. 1**

Gaspar, Georges et al. Marine animal behaviour

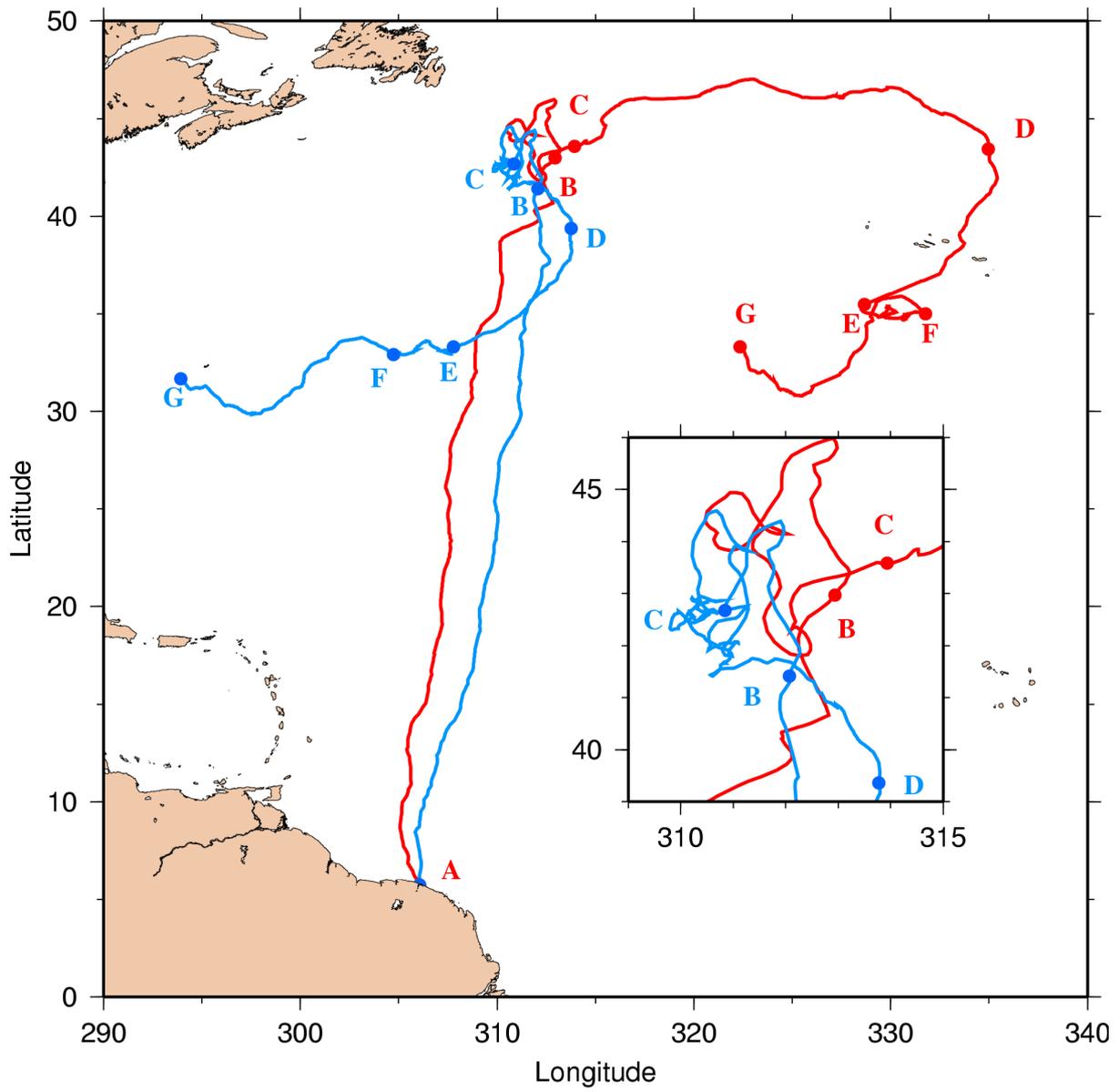

**Fig. 2**

Gaspar, Georges et al. Marine animal behaviour

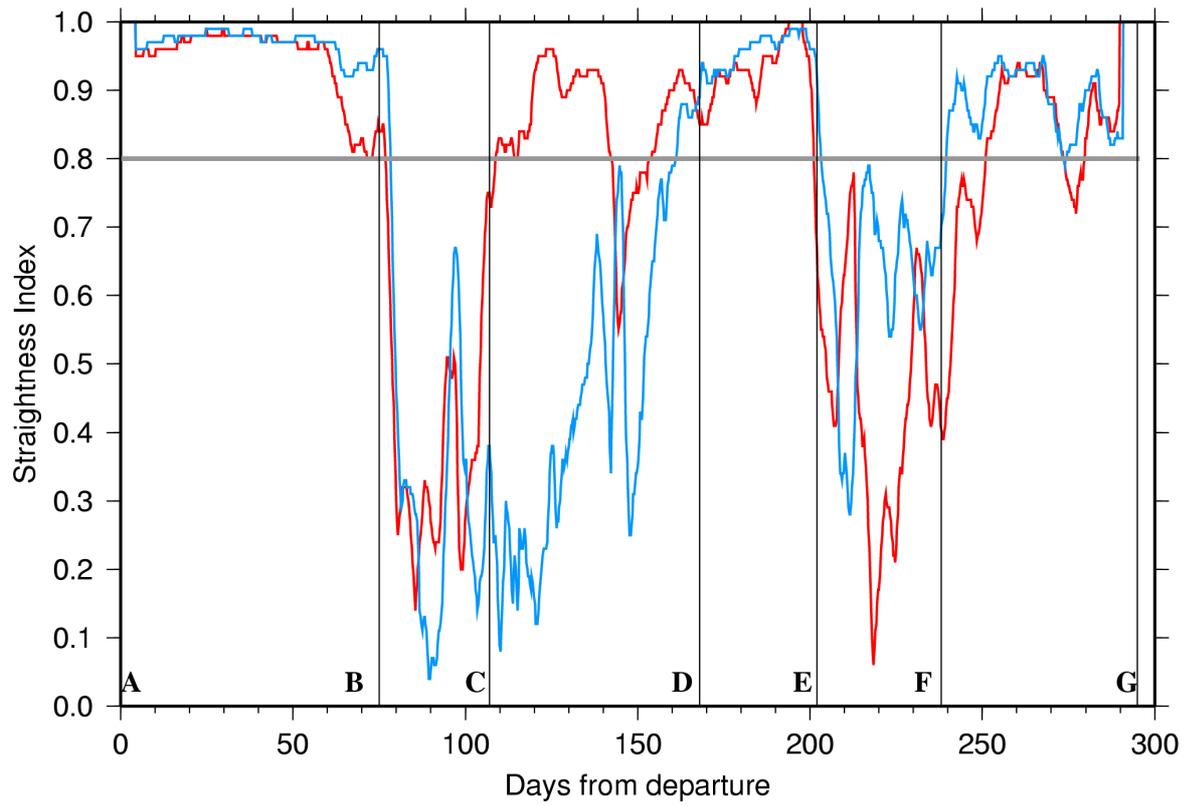

**Fig. 3**

Gaspar, Georges et al. Marine animal behaviour

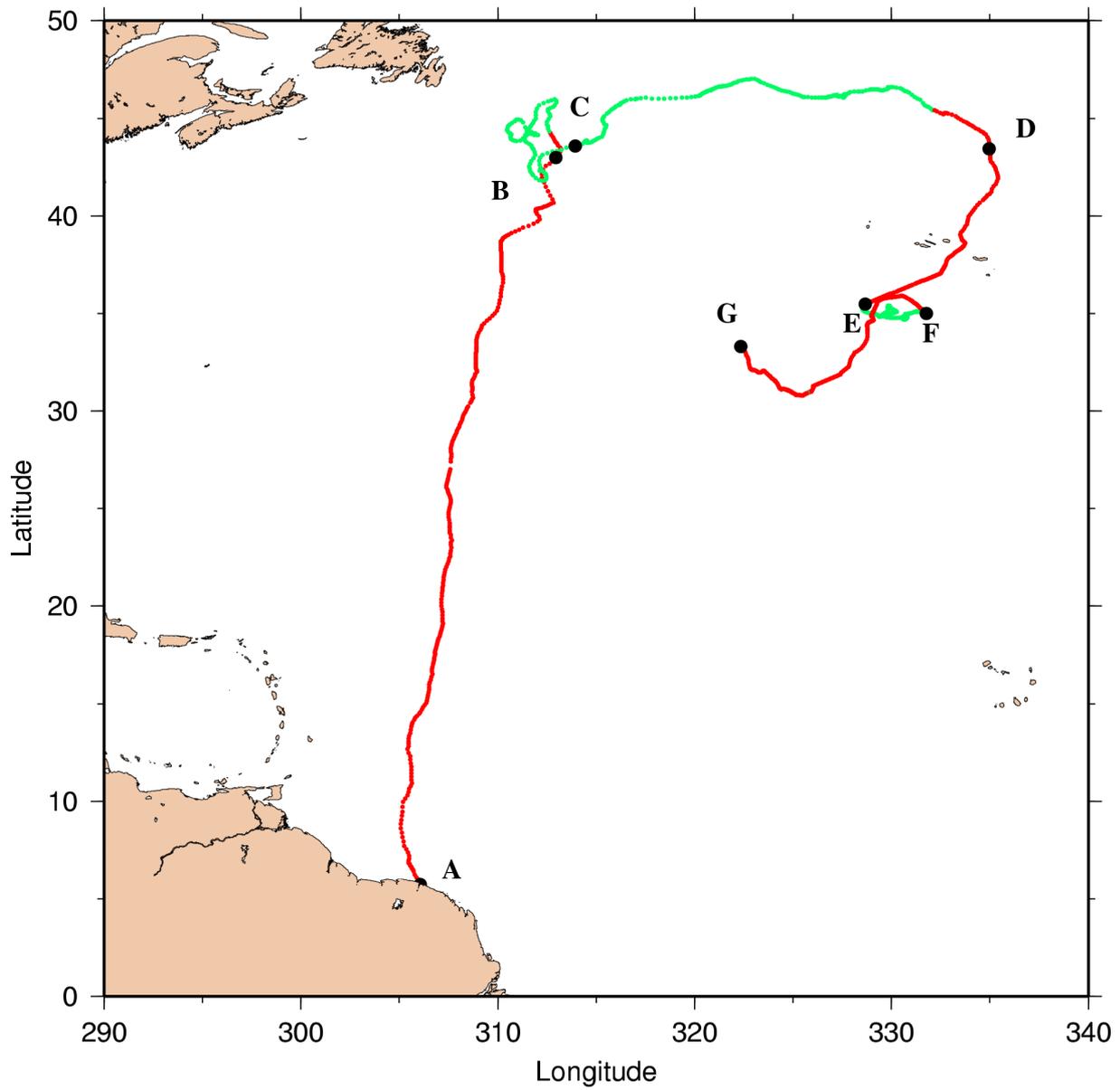

**Fig. 4**

Gaspar, Georges et al. Marine animal behaviour